\documentclass{elsart}

 \usepackage{graphicx}

\usepackage{amssymb}

\begin{document}

\begin{frontmatter}



\title{A Backscatter-Suppressed Beta Spectrometer for Neutron Decay Studies}


\author{F.~E.~Wietfeldt, C.~Trull}
\address{Department of Physics, Tulane University, New Orleans, LA 70118}

\author{R.~Anderman}
\address{Physics Department, Hamilton College, Clinton, NY 13323}

\author{F.~B.~Bateman, M.~S.~Dewey,}
\author{A.~Komives\thanksref{AK}}
\author{\hspace{-0.3em}, A.~K.~Thompson}
\address{National Institute of Standards and Technology, Gaithersburg, MD, 20899}

\author{S.~Balashov\thanksref{SB}}
\author{\hspace{-0.3em}, Yu.~Mostovoy}
\address{Kurchatov Institute, Moscow, Russia}

\thanks[AK]{Present address: Physics Department, DePauw University, Greencastle, IN 46135}
\thanks[SB]{Present address: Particle Physics Department, Rutherford Appleton Laboratory, Oxon, UK}

\begin{abstract}
We describe a beta electron spectrometer for use in an upcoming experiment that will measure the beta-antineutrino correlation coefficient ($a$ coefficient) in neutron beta decay. Electron energy is measured by a thick plastic scintillator detector. A conical array of plastic scintillator veto detectors is used to suppress events where the electron backscattered. A Monte Carlo simulation of this device in the configuration of the $a$ coefficient experiment is presented. The design, construction, and testing of a full-scale prototype device is described. We discuss the performance of this spectrometer with respect to its suitability for the experiment.
\end{abstract}

\end{frontmatter}

\section{Introduction}
\label{}
This paper describes a beta electron spectrometer that is designed for use in an upcoming experiment that will measure the beta-antineutrino correlation coefficient ($a$ coefficient) in neutron beta decay from using a cold neutron 
beam \cite{Yer93,Bal94,Wie04}. The parameter $a$ is a measure of the average correlation in momentum direction of the electron and antineutrino in the decay final state. It can be used, along with other neutron decay observables, to determine the weak vector and axial vector coupling  constants $g_V$ and $g_A$, the first element of the Cabbibo-Kobayashi-Maskawa (CKM) mixing matrix, and set limits on scalar and tensor weak forces, second-class currents, and other possible new forces.
\par
An important systematic effect in the $a$ coefficient experiment arises if the beta electron energy is misidentified. This can happen, for example, if the electron scatters back out of the detector without depositing its full energy. The key performance requirement of this spectrometer is the suppression of such backscattered events with an efficiency of about 90\%. We will describe a novel method of backscatter suppression that is tailored to the specific needs of the $a$ coefficient experiment, a Monte Carlo simulation of this method that tests its performance in the experimental configuration, and the construction and tests of a prototype device using a monoenergetic electron beam. We will show that this spectrometer meets the requirements of the experiment. 
\par
While the spectrometer has been designed according to the needs of a particular experiment, we believe it may be useful in other neutron decay or nuclear beta decay experiments.

\section{Design Criteria and Features}
In the $a$ coefficient experiment beta electrons and recoil protons from neutron beta decay are detected in coincidence. The electron energy, and time-of-flight (TOF) between electron and proton detection, are measured. For each electron energy the protons fall into two distinct TOF groups, fast and slow. The ratio of events in these two groups measures the $a$ coefficient. This technique is described in detail in \cite{Wie04}. If a beta electron with relatively high energy (300--782 keV) backscatters from the energy detector, it will be misidentified as a low energy electron and may fall into the wrong TOF group, generating a systematic error in the measured $a$ coefficient. It is these events that we wish to identify and reject with the beta spectrometer. Events associated with low energy betas, below 300 keV, will remain in the same TOF group even if they backscatter, so backscatter is not a serious problem for them. Our Monte Carlo analysis has shown that, when the beta initial energy is above 300 keV, the fraction of events where the beta backscatters and deposits less than 75\% of the full energy in the energy detector must be no more than 0.5\% to keep the systematic error tolerably small. With a plastic scintillator energy detector and no backscatter suppression, this fraction is expected to be 4--5 times larger than this, so a backscatter suppression scheme is needed. Important measurements of electron backscatter from plastic scintillators by Goldin and Yerozolimsky\cite{Gol04} defined this problem and characterized the size and shape of the backscatter tail in the electron energy response function.
\par
\begin{figure}
\begin{center}
\includegraphics[width= 5.0in]{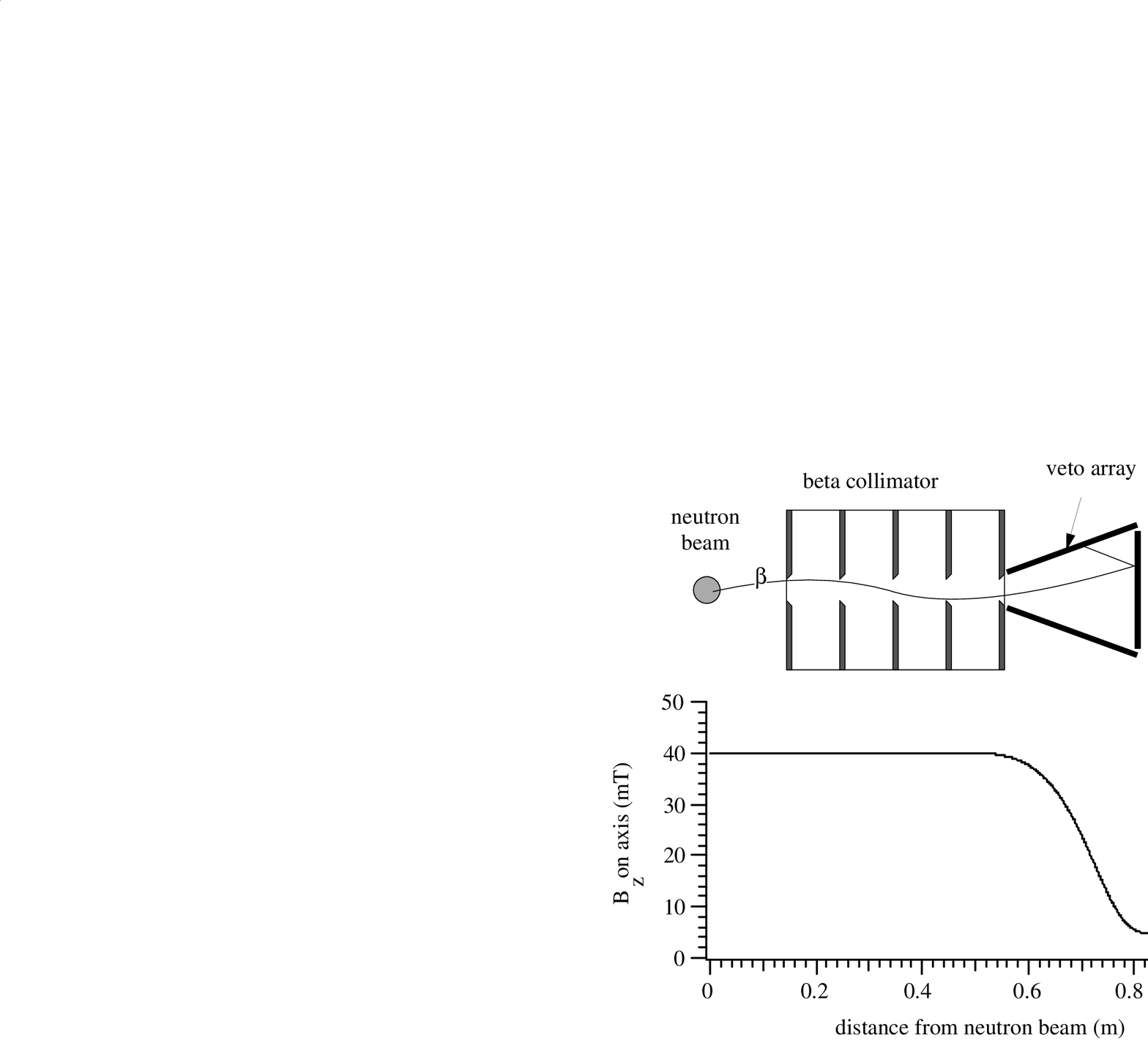}
\end{center}
\caption{\label{F:scheme}Conceptual scheme for a backscatter-suppressed beta spectrometer optimized for 
the planned neutron decay $a$ coefficient experiment. The magnetic field shown is calculated from a realistic
array of coils surrounding the apparatus.}
\end{figure}

The experiment selects beta electrons up to a maximum transverse momentum of about 300 keV/c using a 0.04 T axial magnetic field and a series of circular apertures in the configuration shown in Figure \ref{F:scheme}. The lowest energy betas ($< 350$ keV) are particularly important for the experiment. This represents a large volume of momentum space for low energy electrons that must be accepted with high efficiency by the spectrometer. The spectrometer must also be fairly compact. The experiment will be mounted on a vertical axis, with the beta spectrometer at the bottom, and the spectrometer must fit  in the 72 cm available vertical space between the beta collimator exit and the cold neutron guide hall (experimental hall) floor. 
\par
The energy calibration of the energy detector must be known to better than 1\%. The energy resolution can be much worse than this, provided the energy response function is well understood. This can be accomplished in a separate calibration run using a monochromatic, variable energy electron beam.
\par
The design criteria of the spectrometer can be summarized as follows. It should:
\begin{enumerate}
\item accept a large momentum space volume of low energy ($< 350$ keV) electrons, transported by a weak magnetic field (0.04 T), with high efficiency and measure their energy.
\item identify and reject events where the electron backscattered and the measured beta energy was less than 75\% of its initial energy. If the initial energy was greater than 300 keV the rejection must be effective enough so that less than 0.5\% of these events remain in the data.
\item have a total length of 72 cm or less.
\item be capable of a 1\% accuracy energy calibration.
\end{enumerate}
\par
Identification and rejection of electron backscattering has a long history in beta spectroscopy. A variety of techniques that have been used successfully in previous experiments are not suitable here. Magnetic and electrostatic spectrometers can measure beta energy without backscatter but necessarily have very narrow momentum acceptance. A gas ionization detector could make a backscatter-free measurement, but a detector thick enough to fully stop neutron decay betas requires a window that would cause energy loss issues as bad as or worse than backscattering. Magnetic mirrors can be used \cite{Jack,Mor93} to reflect backscattered electrons back into the energy detector, but are not effective with the weak magnetic field needed for the $a$ coefficient experiment; most betas are not in the adiabatic regime with this field. A thin delta-E detector in front of the energy detector could in principle identify backscattered electrons as they would pass through it twice, but we have found that such a system is not sufficiently discriminating for electrons in the desired energy range. 

\par
We developed a scheme that does satisfy our criteria. The basic idea is illustrated in Figure \ref{F:scheme}. A magnetic field, created by an array of coils that surrounds the experiment, is used to transport beta electrons from the neutron beam, where the decay occurs, to the detector through a series of apertures which we call the beta collimator. In this way electrons with low transverse momentum are preferentially accepted, as desired for the $a$ coefficient experiment. The magnetic field near the axis is 0.04 T, uniform within a 2-cm radius from the axis. The energy detector consists of a circular slab of plastic scintillator located some distance beyond the end of the collimator. It is surrounded by a conical veto array. 
\par
A neutron decay beta electron which is within the momentum acceptance of the collimator/magnetic field arrangement will be transported through the collimator and into the detector chamber. The magnetic field drops rapidly within it. This causes the electron's trajectory to diverge from the axis, and also some of its transverse kinetic energy to be converted into axial energy. Almost all electrons that are accepted by the collimator and enter the detector chamber will strike the energy detector first. If the electron backscatters, it is unlikely that the electron will pass back through the entrance of the detector without first striking the veto array. We note that some backscattered electrons will be reflected back to the energy detector by the magnetic mirror effect, but because most are not in the adiabatic regime, this fraction is negligible.

\section{Monte Carlo Simulation}
\label{S:Monte}
A Monte Carlo simulation of the spectrometer concept shown in Figure \ref{F:scheme} was developed in order to optimize the design and assess its performance. The veto array was modeled as a thick conical plastic scintillator detector
and the energy detector by a circular thick slab plastic scintillator detector. To simplify the problem the detectors were assumed to have perfect (delta function) energy resolution. The effect of a realistic energy resolution on the backscatter suppression efficiency will be investigated using a prototype (see Section \ref{S:proto}). The optimal length and diameters of the veto detector were determined by running the simulation many times and varying those parameters to find the best performance. In the optimal geometry the veto detector had a front (entrance) diameter of 4.0 cm, a rear diameter of 24.0 cm, and a length of 24.0 cm. The energy detector was positioned exactly at the end of the veto detector and had a diameter of 24.0 cm.
\par
In this simulation it was important to account for electron backscatter from a plastic scintillator detector in a realistic way.  To do this we developed a model based on electron backscatter calculations made using the ETRAN \cite{Sel91} electron transport code. In these calculations, electrons with kinetic energy from 100--700 keV, in 100 keV intervals, were directed onto a thick slab of plastic scintillator at seven different incident angles: 0 (normal incidence), 15, 30, 45, 60, 75 and 89 degrees. One million electrons were run for each energy-angle combination. The distributions of backscattered electrons were then fit to phenomenological functions that became the basis of our model.
\par
\begin{figure}
\begin{center}
\includegraphics[width=5.0in]{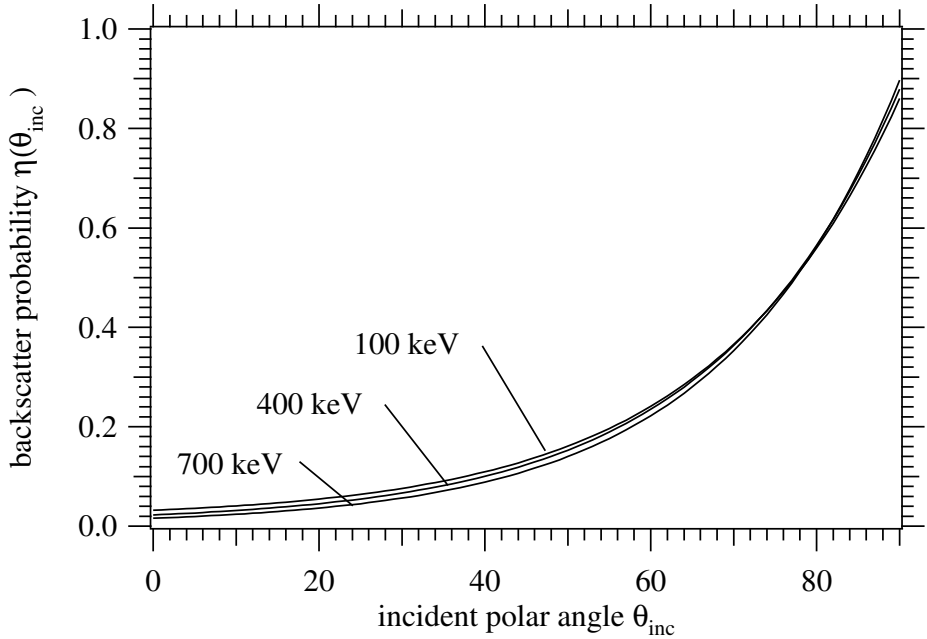}
\end{center}
\caption{\label{F:BSProb} The model probability for electron backscattering as a function of incident energy and angle,
calculated from Equation \ref{E:BSProb}.}
\end{figure}

The total backscatter probability as a function of incident polar angle $\theta_{\rm inc}$ and incident energy was modeled by the function:
\begin{equation}
\label{E:BSProb}
\eta(\theta_{\rm inc}) = A_0 + A_1 \exp (A_2 \, \theta_{\rm inc})
\end{equation}
with $\theta_{\rm inc}$ in degrees and the energy-dependent coefficients $A_0$, $A_1$, $A_2$ given in Table \ref{T:BSProb}. At normal
incidence the backscatter probability is only about 2--3\% but becomes much larger at large angles. The probability decreases slightly with higher incident energy. A plot of this distribution is shown in Figure \ref{F:BSProb}.

\begin{table}
\caption{ \label{T:BSProb} Energy-dependent values of the $A$ coefficients used in Equation \ref{E:BSProb}.}
\begin{tabular}{clll} \\ \hline
Incident energy (keV) & $A_0$ & $A_1$ & $A_2$ \\ \hline
100 & 0.0159 & 0.0160 & 0.0440 \\
200 & 0.0123 & 0.0158 & 0.0443 \\
300 & 0.0088 & 0.0153 & 0.0448 \\
400 & 0.0072 & 0.0156 & 0.0447 \\
500 & 0.0054 & 0.0142 & 0.0459 \\
600 & 0.0045 & 0.0139 & 0.0462 \\
700 & 0.0027 & 0.0132 & 0.0468 \\ \hline
\end{tabular}\\
\end{table}

\par
\begin{figure}
\begin{center}
\includegraphics[width=5.0in]{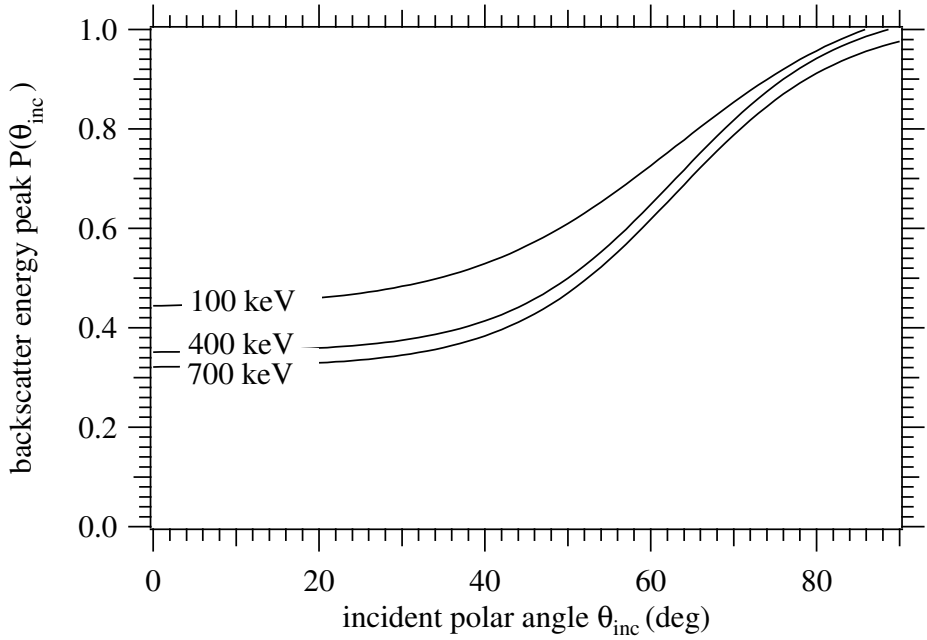}
\end{center}
\caption{\label{F:BSPeak} The incident energy and angle dependent position of the peak in the model backscattered
electron energy distribution function (see Figure \ref{F:BSEnergy}).}
\end{figure}
\begin{figure}
\begin{center}
\includegraphics[width= 5.0in]{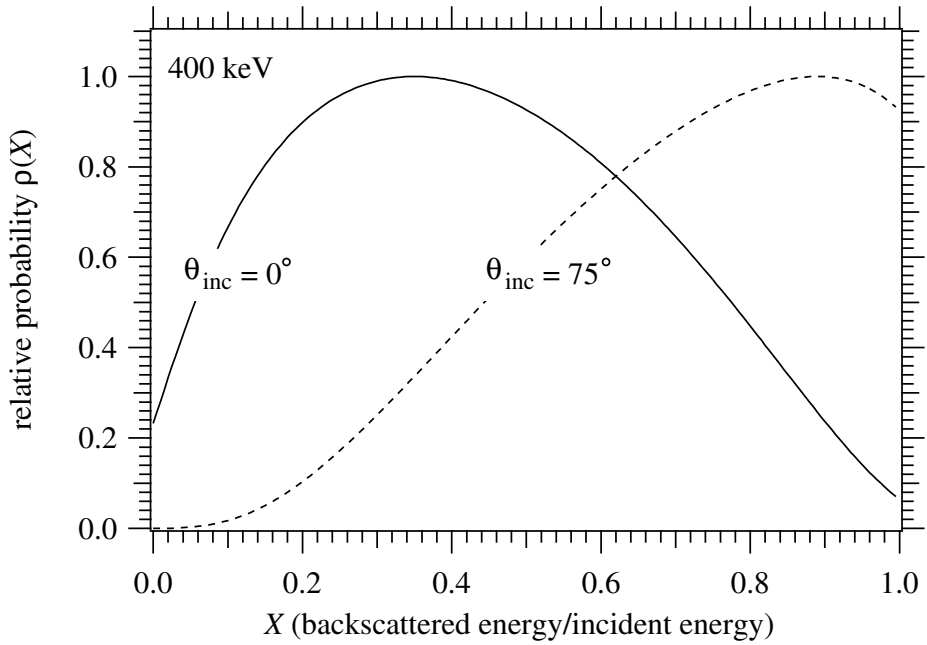}
\end{center}
\caption{\label{F:BSEnergy} The model probability distribution  $\rho(X)$, where $X$ is the fraction of incident energy carried off by the backscattered electron, for the case of a 400 keV electron with normal incidence and $\theta_{\rm inc} = 75^{\circ}$.}
\end{figure}

The energy distribution of backscattered electrons can be fairly well modeled by the following function:
\begin{equation}
\label{E:Rho}
\rho (X) = \exp\left(- \frac{ (X-P)^2 }{ 0.7 \, (0.1 + X) \, (1.2 - X)} \right)
\end{equation}
where $X$ is the ratio of backscattered energy to incident energy. The parameter $P$, which represents the peak
of the distribution, was found to vary with both incident energy and angle. It can be described by:
\begin{equation}
\label{E:BSPeak}
P(\theta_{\rm inc}) = C_0 + C_1 \left(  1 - \frac{1}{ 1 + \exp \left( C_2 ( \theta_{\rm inc} - 63^{\circ}) \right)} \right).
\end{equation}
The coefficients $C_0$, $C_1$, $C_2$ depend on incident energy and are given in Table \ref{T:BSPeak}. A plot of
$P$ as a function of incident angle and energy is shown in Figure \ref{F:BSPeak}. We found that the same functional form (Equation \ref{E:Rho}) provided a good model for all cases; the angle and energy dependence could be accounted for within the single parameter $P$. The backscattered energy fraction
distribution $\rho (X)$, for an incident energy of 400 keV and two different incident angles, 
is shown in Figure \ref{F:BSEnergy}.

\begin{table}
\caption{ \label{T:BSPeak} Energy-dependent values of the $C$ coefficients used in Equation \ref{E:BSPeak}.}
\begin{tabular}{clll} \\ \hline
Incident energy (keV) & $C_0$ & $C_1$ & $C_2$ \\ \hline
100 & 0.44 & 0.65 & 0.08 \\
200 & 0.37 & 0.70 & 0.09 \\
300 & 0.36 & 0.70 & 0.10 \\
400 & 0.35 & 0.70 & 0.10 \\
500 & 0.32 & 0.70 & 0.10 \\
600 & 0.32 & 0.70 & 0.10 \\
700 & 0.32 & 0.70 & 0.10 \\ \hline
\end{tabular}\\
\end{table}

\par
\begin{figure}
\begin{center}
\includegraphics[width= 5.0in]{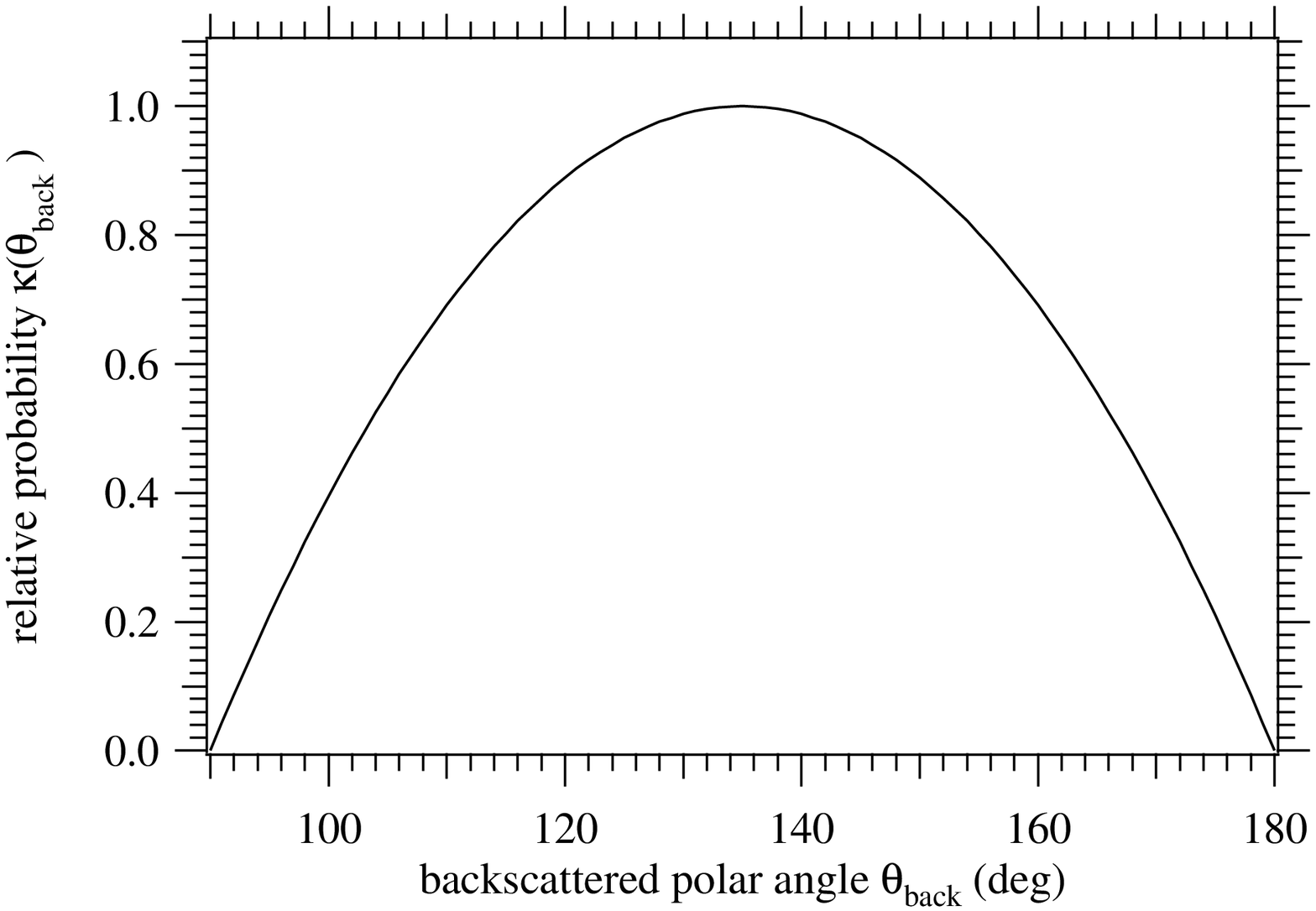}
\end{center}
\caption{\label{F:BSTheta} The model distribution $\kappa$ of backscattered electron polar angle $\theta_{\rm back}$.}
\end{figure}

The distribution of polar angle of backscattered electrons was found to be well-represented by a parabola with a peak
at 135$^{\circ}$:
\begin{equation}
\kappa(\theta_{\rm back}) = 1 - \frac{ ( \theta_{\rm back} - 135^{\circ} )^2 }{2025},
\end{equation}
where $\theta_{\rm back}$ is measured with respect to the incident normal, {\em i.e.} $\theta_{\rm back} = 180^{\circ}$
corresponds to backscatter perpendicular to the surface. This distribution was found to be quite independent of incident energy and angle. Note that an isotropic distribution would follow a sine function peaked at $\theta_{\rm back} = 90^{\circ}$, so this distribution shows that backscatter into wide angles, away from the normal, is somewhat suppressed. Figure \ref{F:BSTheta} shows this distribution.

\par
Finally, the distribution of backscattered azimuth angle is found to be approximately uniform from 0 to 360$^{\circ}$ when
the incident polar angle is small ($\theta_{\rm inc} < 15^{\circ}$). For larger incident angles the distribution is approximately
a Gaussian function centered on the incident azimuthal angle, with a width that decreases with increasing incident
polar angle: 
\begin{equation}
\lambda( \phi_{\rm inc} ) = \exp \left( -\frac{ \left( \phi_{\rm back} - \phi_{\rm inc} \right)^2}
{2 \left( \frac{2500}{\theta_{\rm inc} } \right)^2 } \right) \qquad \mbox{(all angles in degrees).}
\end{equation}
This behavior is reasonable. For large incident polar angles some memory of the incident azimuthal angle should be retained, less so for incident polar angles close to normal.
\par
In the Monte Carlo simulation, each electron began at a random point in the neutron beam region (see Figure \ref{F:scheme}) with its momentum in a random direction. It was initially tested
to see if its axial momentum was toward the beta spectrometer and its transverse momentum would be accepted by the collimator system in a uniform 0.04T axial magnetic field. If it passed this test it was included in the simulation.  The electron was then transported through the geometry in Figure \ref{F:scheme}, using the calculated magnetic field shown. Realistic electron trajectories were calculated by 4th-order Runge-Kutta integration \cite{Gea71} of the Lorentz force law:
\begin{equation}
m \frac{d^2\vec{r}}{dt^2} = -e \vec{v}\times \vec{B}.
\end{equation}
When an electron struck the energy detector or the veto detector, it was either absorbed or backscattered according to the model probability distributions given above. An electron that struck the veto detector was considered ``vetoed'' if at least 50 keV was deposited there. If after first entering the spectrometer an electron backscattered and was later transported back out through the entrance, that electron was considered lost, but any energy it deposited in the detectors was recorded.
\par
The following results were obtained using the optimal geometry for the veto detector described above. One million accepted electrons were transported for each initial kinetic energy. 
A small fraction of electrons ($<$2\%) failed to enter the spectrometer, even though they passed the original momentum test, due to the gradient in the magnetic field near the end of the collimator. These electrons were tallied as {\sf noEnter}.
A tally was also kept of the number of electrons that were fully absorbed in the energy detector ({\sf nAbs}), the number that struck the energy detector first and subsequently backscattered ({\sf nBack}), the number of the these that were vetoed ({\sf nVeto}), the number of backscattered electrons that deposited less than 75\% of the initial energy 
in the energy detector ({\sf nBack$_{\sf 75}$}), the number of the those that were vetoed ({\sf nVeto$_{\sf 75}$}), and finally the number of electrons that struck the veto detector first, regardless of whether the energy detector was later struck ({\sf nV1st}). These results are presented in Table \ref{T:MCResults}.
Also shown in this table are the fraction of electrons in the tail of the energy detector spectrum with $<$75\% of full energy, with no veto ({\sf f$_{\sf 75}$}), and after subtracting vetoed events ({\sf f$_{\sf 75}$V}). The last row shows the veto efficiency of the spectrometer ({\sf vetoEff$_{\sf 75}$}), defined as 1 - ({\sf f$_{\sf 75}$V})/{\sf f$_{\sf 75}$}.

\begin{table}
\caption{ \label{T:MCResults} Results from the Monte Carlo simulation of the spectrometer, using the optimal geometry. 
A total of 1 million electrons were run at each energy.}
\begin{tabular}{llllllll} \\ \hline
Energy (keV) & 100 & 200 & 300 & 400 &  500 & 600 & 700\\ \hline
{\sf noEnter}          & 1851 & 104 & 1 & 0 & 0 & 6 & 18 \\
{\sf nAbs}               & 942991 & 953412 & 963884 & 967593 & 973125 & 975773 & 978899 \\
{\sf nBack}             & 71492 & 46484 & 36115 & 32407 & 26875 & 24261 & 21083 \\
{\sf nVeto}             & 34757 & 33342 & 28928 & 27126 & 22944 & 21243 & 18852 \\
{\sf nBack$_{\sf 75}$}   & 64115 & 37544 & 27715 & 24705 & 19845 & 17754 & 15588 \\
{\sf nVeto$_{\sf 75}$}   & 34757 & 33342 & 25678 & 22809 & 18424 & 16635 & 14761 \\
{\sf nV1st}             & 1660   & 0 & 0 & 0 & 0 & 0 & 0 \\ 
{\sf f$_{\sf 75}$}       &  0.067 & 0.040 & 0.030 & 0.025 & 0.021 & 0.018 & 0.017\\
{\sf f$_{\sf 75}$V} & 0.031 & 0.0044 & 0.0021 & 0.0020 & 0.0015 & 0.0011 & 0.0008\\
{\sf vetoEff$_{\sf 75}$} & 0.54 & 0.89 & 0.93 & 0.92 & 0.93 & 0.94 & 0.95\\ \hline
\end{tabular}\\[1ex]
The rows are defined as follows:
\begin{tabbing}
xxxxxxxx\= = \= \kill
{\sf noEnter}  \> = \> the number of electrons that failed to enter the spectrometer.\\
{\sf nAbs} \> = \> the number of  electrons that deposited full energy in the energy\\
\> \> detector.\\
{\sf nBack} \> = \> the number that backscattered from the energy detector.\\
{\sf nVeto} \> = \> the number of backscattered electrons that were vetoed \\
\> \>(deposited $>$50 keV in the veto detector).\\
{\sf nBack$_{\sf 75}$} \> = \> the number that backscattered and deposited $<$75\% of full energy\\
 \> \>  in the energy detector.\\
{\sf nVeto$_{\sf 75}$} \> = \> the number of nBack75\% that were vetoed.\\
{\sf nV1st} \> = \> the number of electrons that struck the veto detector first.\\
{\sf f$_{\sf 75}$} \> = \> the fraction of electrons in the tail of the energy detector spectrum\\
\> \>  with $<$75\% of full energy, with no veto.\\
{\sf f$_{\sf 75}$V}  \> = \> the fraction of electrons in the tail of the energy detector spectrum\\
\> \>  with $<$75\% of full energy, after subtracting vetoed events.\\
{\sf vetoEff$_{\sf 75}$} \> = \> the veto effficiency for backscattered electrons that deposited\\
\> \>  $<$75\% of full energy in the energy detector.
\end{tabbing}
\vspace{1.0in}
\end{table}

\par
\begin{figure}
\begin{center}
\includegraphics[width= 5.0in]{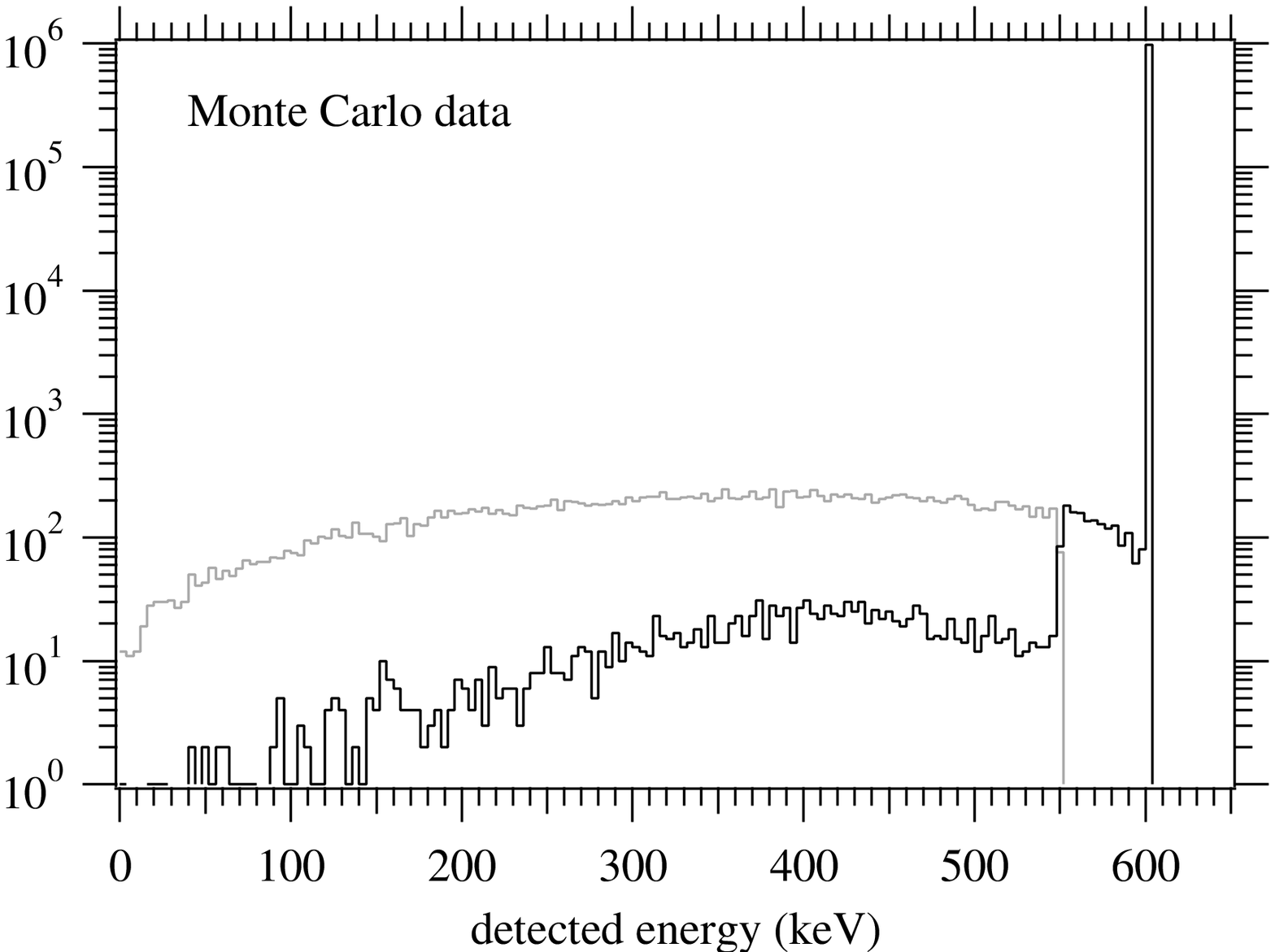}
\end{center}
\caption{\label{F:BS600} Monte Carlo simulation of the energy deposited in the energy detector with (gray) and without (black) a coincidence with the veto detector. The initial electron energy was 600 keV. Events in the energy range 550--600 keV cannot have a coincidence because of the requirement that at least 50 keV be deposited in the veto detector. Non-coincidence events below 550 keV were caused by electrons escaping back through the entrance of the spectrometer without depositing sufficient energy in the veto detector.}
\end{figure}

We note that 94--98\% of all electrons within the desired momentum acceptance were detected and measured by the energy detector in the simulation, so our first design criterion is satisfied. The electrons that failed to enter the 
spectrometer or struck the veto detector first are effectively lost to the experiment, but their number is tolerably small.
The backscatter veto efficiency was relatively poor for 100 keV electrons, mainly because of the requirement that the
electron deposit at least 50 keV in the veto detector to be considered vetoed. In the energy range of importance,
300--700 keV, the veto efficiency was greater than 90\%, and the fraction of events in the low-energy tail of the
energy detector response, with less than 75\% full energy, was about 0.2\% or less, so the second design criterion
was also satisfied in the simulation. Figure \ref{F:BS600} shows the Monte Carlo spectra for energy deposited in the energy detector with and without a coincidence with the veto detector, using an initial electron energy of 600 keV. 

\par
The third design criterion can be achieved. The length of the active elements of the spectrometer in the optimal configuration is 24 cm. With an efficient design the total length, including photomultiplier tubes and front end electronics, will be less than 72 cm. The fourth criterion is no problem using a plastic scintillator. The energy resolution, or width of the energy response function, will be much broader than 1\%, but the centroid and shape can be determined very precisely, with an offline calibration, to achieve a 1\% energy calibration.

\section{Prototype Spectrometer}
\label{S:proto}
The Monte Carlo simulation described in the previous section demonstrated that under ideal conditions this scheme for a backscatter-suppressed detector is successful. However some of the assumptions in the Monte Carlo were simplistic, for example the conjecture that any electron that deposits at least 50 keV in the veto detector will produce a veto signal. In reality the probability of creating a usable veto pulse depends on the scintillation light collection efficiency which is highly position-dependent; it will not be a simple threshold. Also we cannot be certain that the ETRAN-generated backscatter probability distributions are sufficiently realistic. It is essential to show that this scheme will work in practice before undertaking the effort and expense of the $a$ coefficient experiment that will depend on it. Therefore we decided to design, build, and test a prototype spectrometer. In addition to proving the effectiveness of this concept, the process of building the prototype would expose any unforeseen technical challenges in the scheme.
\par
The magnetic field shown in Figure \ref{F:scheme} will be essential to the proper operation of the spectrometer in the $a$ coefficient experiment, but it is not necessary for testing the prototype. It suffices to use a collimated electron source with no magnetic field for the prototpye. In this mode the probability for backscatter from the energy detector is lower because the electrons will be closer to normal incidence (see Figure \ref{F:BSProb}). It is also less likely for a backscattered electron to escape through the spectrometer entrance without striking the veto detector, because a magnetic field near the entrance aids in electron escape. Therefore from the point of view of an ideal apparatus the prototype should perform better, {\em i.e.} produce a smaller unvetoed backscattered tail, than the spectrometer for the $a$ coefficient experiment. However the ``non-idealities'' such as scintillation yield, light collection efficiency, and photomultiplier tube (PMT) performance (assuming they are magnetically shielded) are approximately the same in both. The prototype test serves as a measure of the departure of a real apparatus from the ideal behavior of the Monte Carlo. When considered together, the Monte Carlo simulation and prototype performance provide a reliable picture of how well the spectrometer will perform in the experiment.
\par
\begin{figure}
\begin{center}
\includegraphics[width= 3.5in]{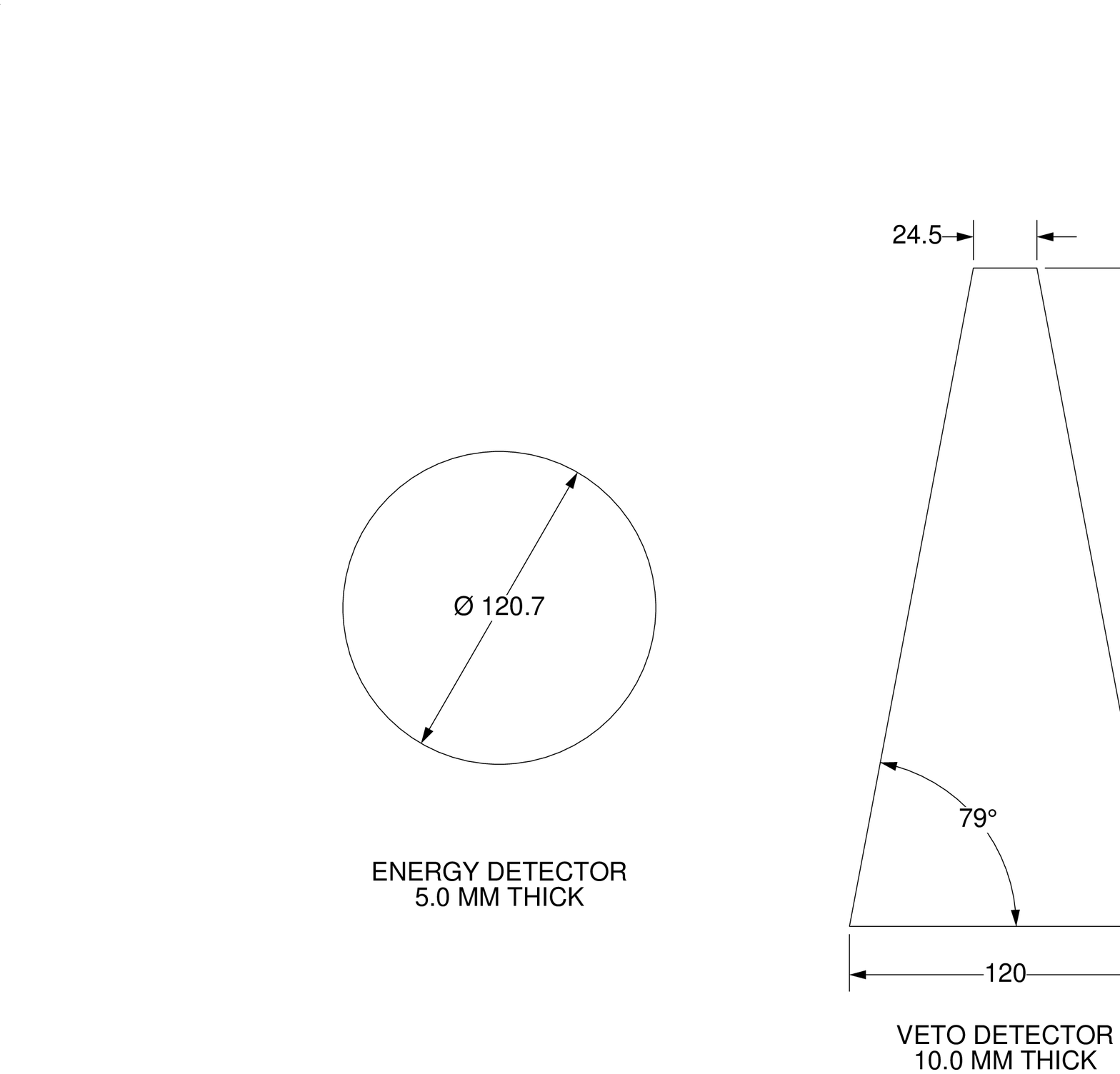}
\end{center}
\caption{\label{F:scintillators} The scintillator element shapes that were used for the energy detector (left) and the veto detector (right). The veto detector consisted of six trapezoidal elements like this arranged into a close-fit hexagonal cone. All dimensions are mm.}
\end{figure}

Bicron BC-408 \cite{NISTdisc} plastic scintillator was used for the energy detector and veto detector. Figure \ref{F:scintillators} shows the scintillator elements. The energy detector was a circular slab, 120.7 mm in diameter and 5.0 mm thick. The veto detector consisted of six trapezoidal slabs, each 10 mm thick,  fit closely together (less than 1 mm separation) along their lateral edges to form a hexagonal cone with inner diameter 45 mm in the front and 210 mm in the back, as measured by their inscribed circles. The light guides were fabricated from Polycast UVT acrylic and polished to a mirror finish using micron sapphire polish. The energy detector was glued to a cylindrical light guide of equal diameter, and each veto paddle (trapezoidal segment) was edge-coupled to a light guide that makes a 79$^{\circ}$ bend. The light guides were designed to be nearly adiabatic, {\em i.e.} the cross sectional area was kept constant or gradually increased from the scintillator end to the PMT end, for maximum light collection. The veto paddle configuration is shown in Figure \ref{F:vPaddle}. Optical cement that matched the index of refraction of the scintillator and light guides was used for all optical joints.

\begin{figure}
\begin{center}
\vspace{1in}
SEE FIGURE AT END OF FILE\\[1in]
\end{center}
\caption{\label{F:vPaddle} An assembled veto paddle including the light guide. Six of these were used to form the veto detector.
All dimensions are mm.}
\end{figure}
\par
The detectors were installed in a vacuum chamber as shown in Figure \ref{F: prototype}. The chamber was stainless steel with welded construction and flanges were aluminum with Viton o-ring seals. All seven light guides penetrated the vacuum chamber through Viton bayonet compression seals. On the inside of the chamber, the scintillators and light guides were wrapped with aluminized Mylar foil for optical isolation. Outside the vacuum chamber, each light guide was coupled to a PMT using optical grease. Five of the veto detectors used Burle 8575 2-inch PMT's and the sixth used a Burle 8850 2-inch PMT. 
The latter has a high-gain first dynode that is optimized for few-photon counting. This was done in order to compare the performance of these two PMT's in this application. A single Burle 8854 5-inch PMT was used for the energy detector. A thin Kapton film, 25 $\mu$m thick, was used for the vacuum window at the entrance of the spectrometer.
\begin{figure}
\begin{center}
\vspace{1in}
SEE FIGURE AT END OF FILE\\[1in]
\end{center}
\caption{\label{F: prototype} The energy and veto detectors, with light guides, arranged inside the vacuum chamber. The photomultiplier tubes are not shown. All dimensions are mm.}
\end{figure}
\par
It may seem that the 79$^{\circ}$ bending angle of the veto paddle light guides is excessive. Such a large bending angle reduces the light collection efficiency of the guides. We did this for the reason that, in the spectrometer to be used for the $a$ coefficient experiment, the energy detector will need to be much larger (at least 240 mm diameter) than that of the prototype. A more complicated system of light guides and PMT's will be needed for good light collection with the larger detector. The veto detector PMT's must attach to the chamber without mechanical interference with the energy detector. The extreme case is for the veto detector PMT's to be oriented perpendicular to the axis of the chamber; this maximizes the lateral space available for the energy detector. This is the ``worst case scenario'' for the veto detector light guides, and makes a good test case for the prototype. In the final version of the spectrometer it should be possible to relax this constraint and orient the veto detector PMT's at a smaller angle to the chamber axis, in which case the bending angle of the light guides will be less and their light collection efficiency improved. 
\par
High voltage was distributed to the PMT's from a computer controlled 8-channel 0--5000 V (negative) power supply. High voltage was applied to the photocathode of each PMT, with the anode at ground. Resistive divider circuits recommended by the PMT manufacturer were used to apply voltage to the dynodes. 

\section{Electron Beam Tests}
The prototype spectrometer was tested at the NIST Van de Graaff accelerator facility using a nominal electron beam energy of 1 MeV and very low current. 
The Van de Graaff accelerator is a direct-current machine supplying electron beams in the energy range of approximately 0.5 MeV to 2.5 MeV at currents from below 1 pA up to about 200 $\mu$A. Electron beams are emitted from a pentode-type heated-cathode emission circuit and accelerated down a vertical acceleration tube.  After acceleration, beams may be extracted from a straight-through vertical port or steered toward one of two horizontal beam ports via two 45-degree bending magnets.   For these tests, the 45-degree horizontal port was chosen to yield the best energy definition and reduce the low-energy component due to scattered electrons striking the walls of the beam pipe.   Beam conditions were monitored using a solid-state Si(Li) detector placed in front of the spectrometer prior to the calibrations and at periodic intervals during the experiment.  A typical beam spectrum is depicted in Figure \ref{F:VdGbeam}. As seen in this figure, the total energy spread is about 4.5\%.  Accounting for a detector resolution of about 3\%, the energy spread of the beam is about 3.4\%. The low energy ``tail'' in this spectrum is mostly due to electrons that backscattered from the Si(Li) detector, but some can be attributed to scattered electrons transmitted by the system, and possibly a small contribution from induced x rays.
\begin{figure}
\begin{center}
\vspace{1in}
SEE FIGURE AT END OF FILE\\[1in]
\end{center}
\vspace{-1cm}
\caption{\label{F:VdGbeam} Energy spectrum of the 1 MeV electron beam produced at the NIST Van de Graaff accelerator, measured with a Si(Li) detector.}
\end{figure}
 \par
\begin{figure}
\begin{center}
\includegraphics[width=5.0in]{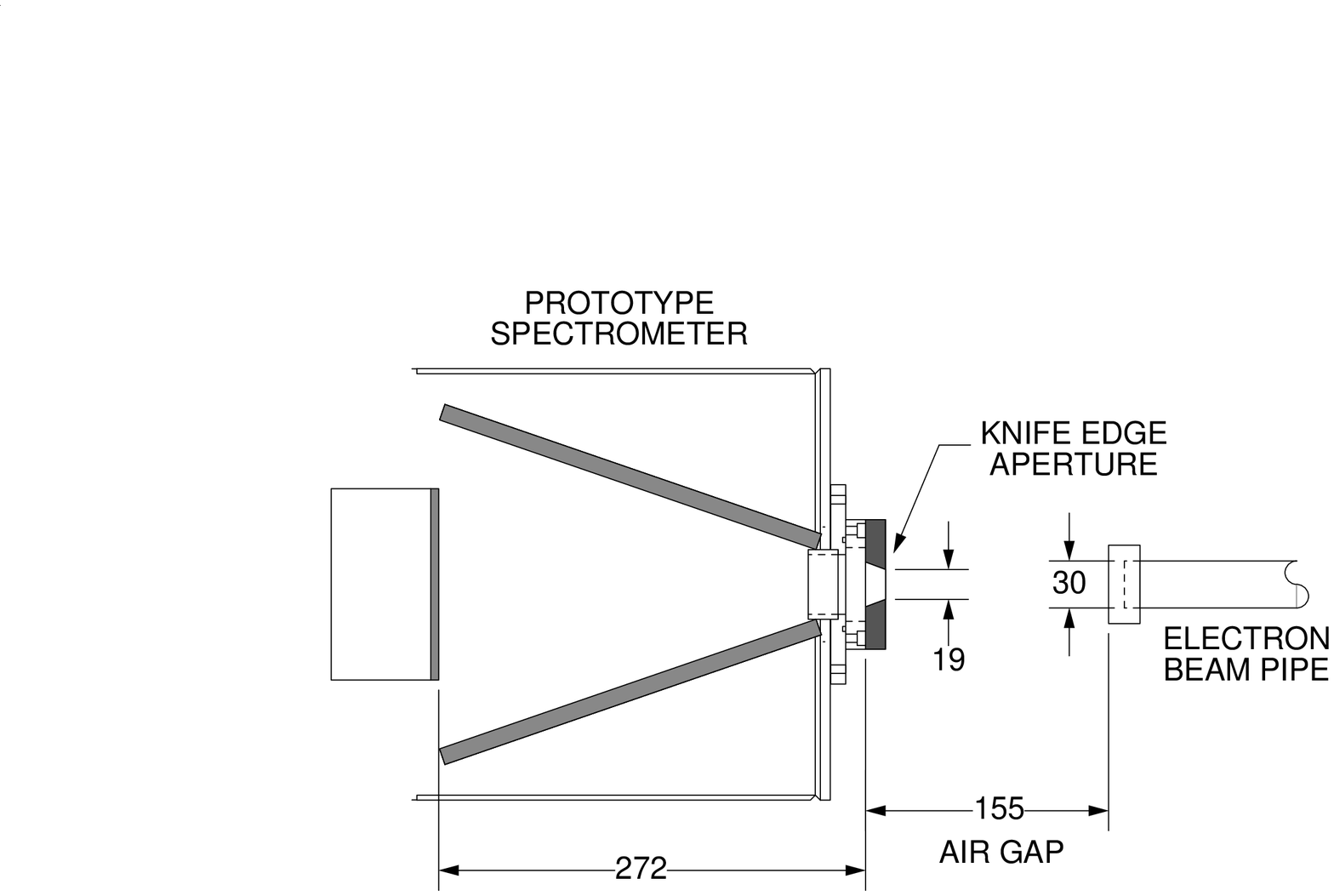}
\end{center}
\caption{\label{F:beamSet} Arrangement of the prototype spectrometer on the NIST Van de Graaff electron beam. All dimensions in mm.}
\end{figure}

The entire prototype assembly was transported to the Van de Graaff facility, set into position and aligned on the 45-degree beam port as depicted in Figure \ref{F:beamSet}.  The beam port vacuum window was 25 $\mu$m thick Kapton film, and a 155-mm air gap separated it from the spectometer window. A lead (Pb) knife-edge aperture was used to collimate the beam and minimize the possibility of scattered electrons entering the spectrometer. A rough vacuum of approximately 0.1 torr was maintained inside the spectrometer during the beam tests. The beam current was maintained at less than 100 nA to avoid overwhelming the data acquisition system.  After subtacting the expected energy loss in the two Kapton windows and the air gap between them the electron kinetic energy incident on the energy detector was calculated to be 976 keV.
\par
\begin{figure}
\begin{center}
\includegraphics[width=5.0in]{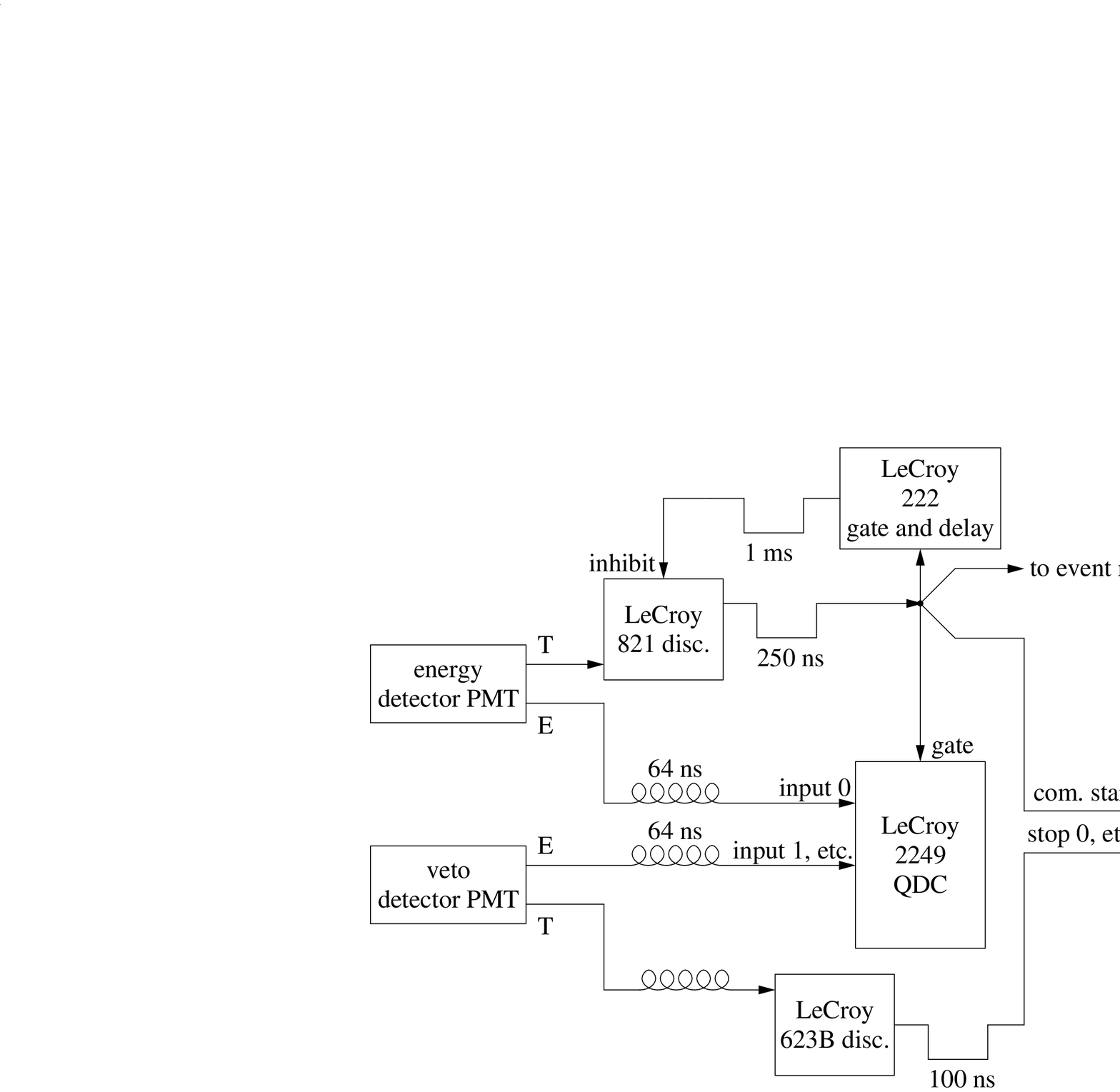}
\end{center}
\caption{\label{F:DAQ} The data acquisition system for the Van de Graaff test run. Only one of six veto channels is shown, the other five are similar.}
\end{figure}

The data acquisition system is depicted in Figure \ref{F:DAQ}. It is a NIM/CAMAC system controlled by a desktop PC. The anode signal from each PMT is split into identical timing (T) and energy (E) signals. The energy detector timing signal is sent to a level discriminator that produces the event trigger. The trigger gates the CAMAC charge-integrating analog-to-digital converter (QDC),  sends a common start to the CAMAC time-to-digital converter (TDC), and generates a gate that inhibits another event for 1 ms, enough time for all conversions and readout. The energy detector and six veto detector energy signals are delayed by 64 ns so they fall comfortably within the QDC gate, and distributed to 7 QDC inputs. The timing signal from each veto PMT is delayed and sent to another level discriminator which produces a TDC stop signal for each channel. The energy detector and all veto detector QDC outputs, and all TDC outputs, are recorded for each event.

\section{Results and Analysis}

\begin{figure}
\begin{center}
\includegraphics[width= 5.0in]{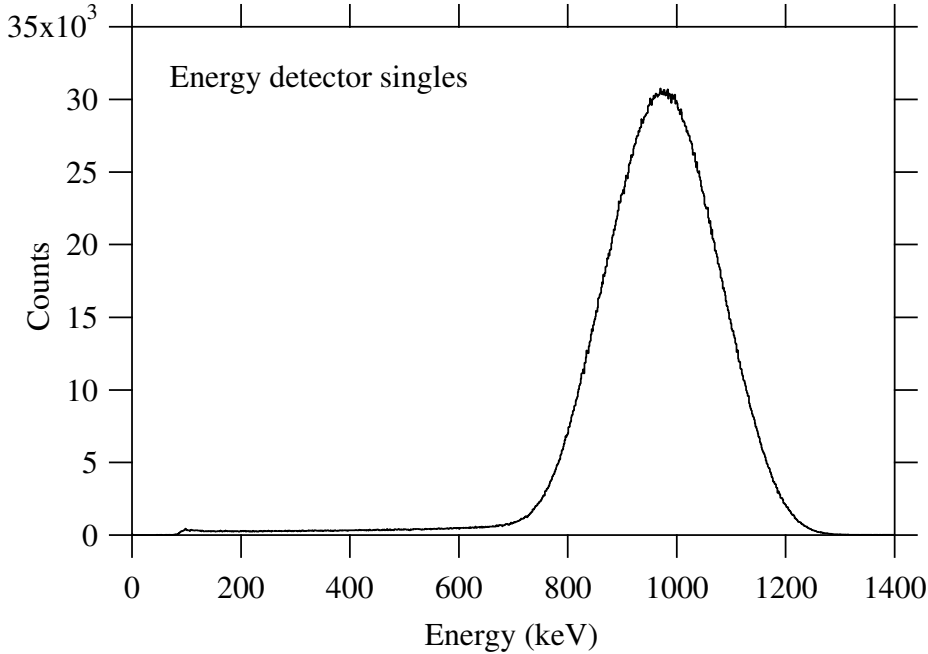}
\end{center}
\caption{\label{F:eDetLin} Energy detector singles events recorded using the 976 keV electron test beam.}
\end{figure}

The analyzed data set represents 7200 s (real time) of beam at an average electron
rate of about 830 s$^{-1}$. In these data the energy detector and veto detector PMT base voltages were 1770 V and 2900 V, respectively.
Figure \ref{F:eDetLin} shows the energy detector singles
spectrum, {\em i.e.} irrespective of the signal in the veto detector. The energy resolution is 26\%
FWHM. The non-Gaussian low-energy tail of the spectrum is about 4\% of the total.
\par
\begin{figure}
\begin{center}
\includegraphics[width= 5.0in]{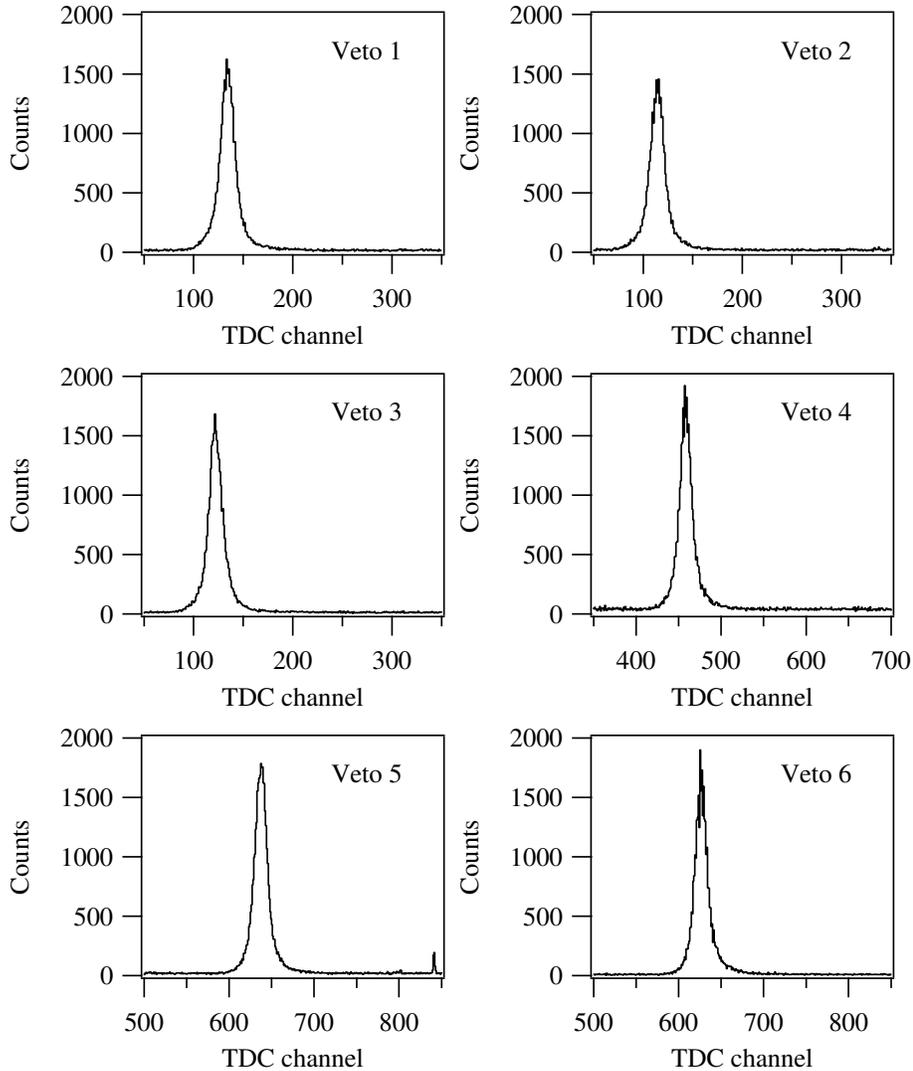}
\end{center}
\caption{\label{F:TDCspec} Timing spectra from the test run. The TDC start was generated by the energy detector and the stop by each of six veto detectors. One TDC channel equals 4.3 ns.}
\end{figure}
\begin{figure}
\begin{center}
\includegraphics[width= 5.0in]{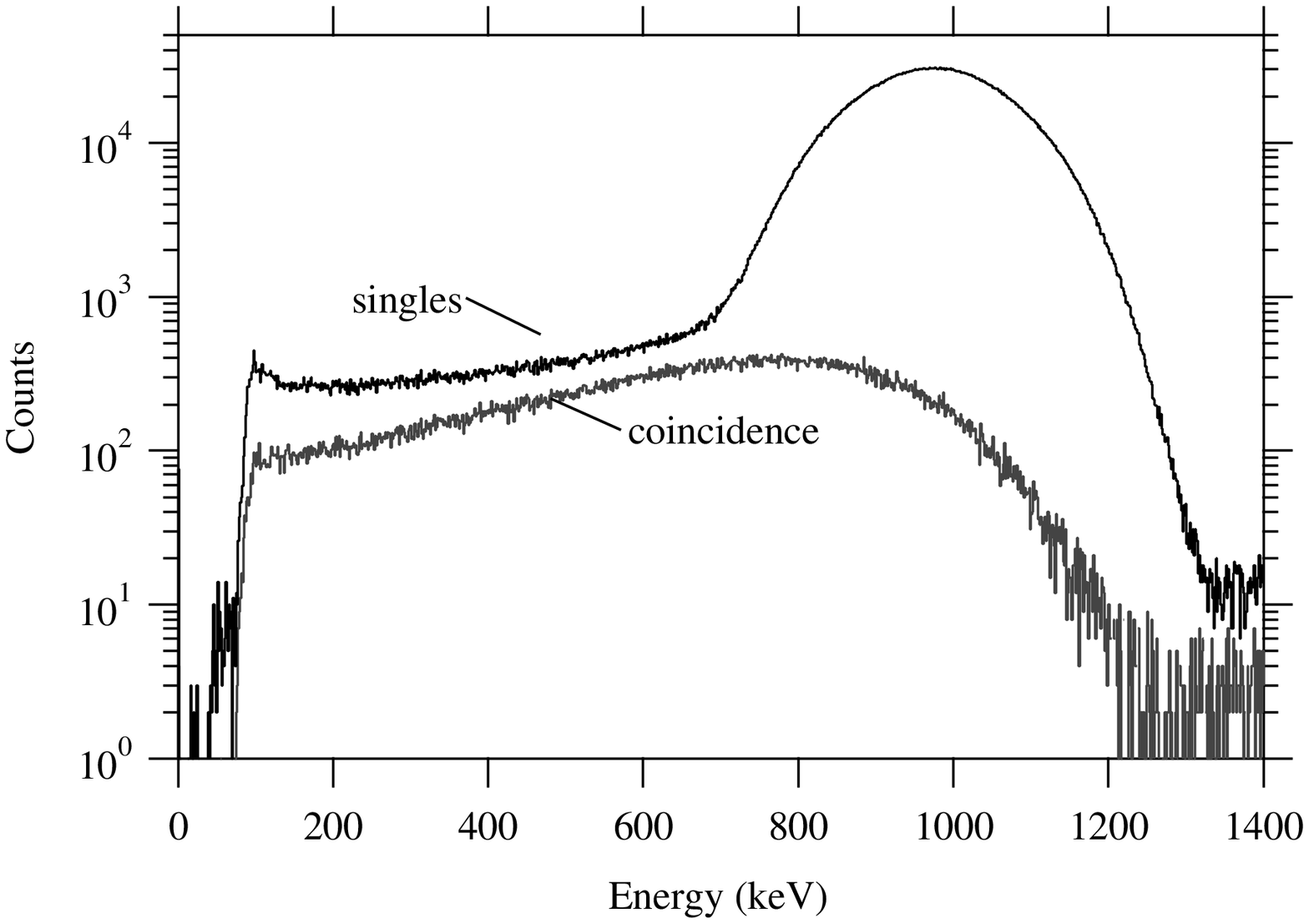}
\end{center}
\caption{\label{F:eDetLog} Energy detector singles events, and coincidence events with one or more veto detectors.}
\end{figure}

Figure \ref{F:TDCspec} shows the relative timing between the energy detector (start)
and each veto paddle (stop) for coincidence events as measured by the TDC. Each peak represents true coincidences, primarily electrons that backscattered from the energy detector into the
veto paddle, although other types of coincidence events are possible (see discussion below). Differences in the peak positions are due to different amounts of cable delay used for the veto paddles. The heights and widths
of the six peaks are comparable which indicates good uniformity in the performance of the six
veto paddles. An energy-veto coincidence spectrum for the data set was generated by including
only events which fell inside a 100-channel-wide window around the coincidence peak in at least one of the six TDC spectra. Random coincidences were subtracted by generating a second, random coincidence spectrum that required the event to fall inside either of two 50-channel-wide windows adjacent to (above and below), but not including, the peak in at least one TDC spectrum. Random coincidences are, to good approximation, distributed linearly in each TDC spectrum so this accounts for random coincidences in the peak region quite well. The second coincidence spectrum was subtracted from the first to produce a true coincidence spectrum. This is shown in Figure \ref{F:eDetLog} along with the energy singles spectrum (the same as in Figure \ref{F:TDCspec} except now on a log scale). The net coincidence spectrum contains 2.8\% of the singles spectrum and it accounts for about 70\% of its low energy tail.

\begin{figure}
\begin{center}
\includegraphics[width=5.0in]{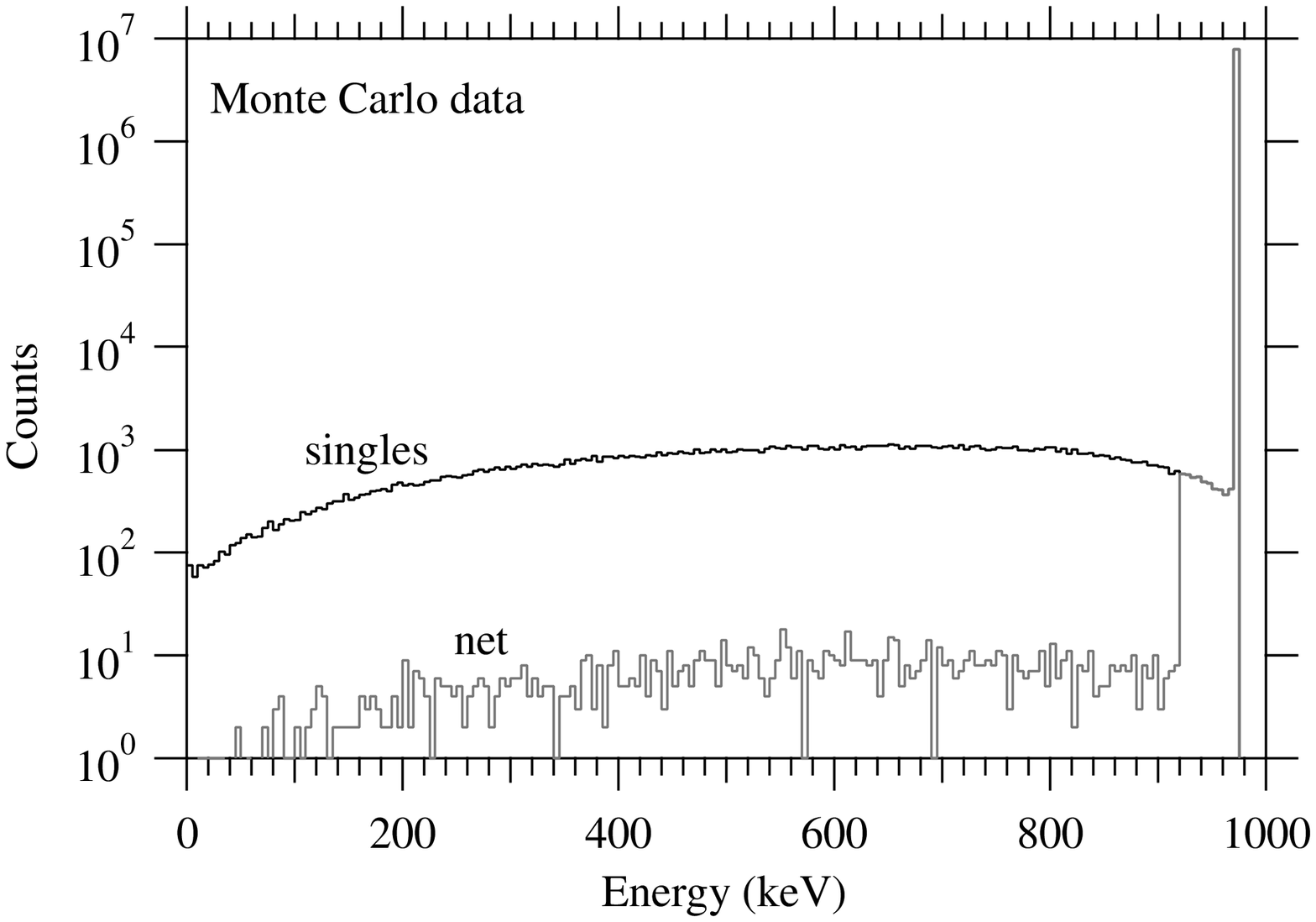}
\end{center}
\caption{\label{F:MCproto} Monte Carlo data using the conditions of the prototype test run, using an electron energy of 976 keV. Energy detector singles events, and net events (singles minus coincidences) are shown. The net events are those that were not vetoed.}
\end{figure}

In order to help interpret these results we developed a second set of Monte Carlo data. This set used the same algorithm and code as described in Section \ref{S:Monte} except now the electron energy (976 keV) and beam collimation simulated the prototype test run instead of the $a$ coefficient experiment. The magnetic field in this simulation was set to zero. Figure \ref{F:MCproto} shows the results of this simulation. The low energy tail in the singles spectrum contains 1.8\% of the total events and this represents the total fraction of electrons that backscatter from the energy detector.
Note that there are no coincidences above 926 keV due to the artificial requirement that at least
50 keV be deposited in the veto detector to produce a coincidence. In the region 0--925 keV the integrated coincidence spectrum equals 99.2\% of the integrated singles, indicating that 99.2\% of
backscattered electrons were vetoed. The remainder are electrons that backscattered and exited through the entrance of the spectrometer without striking the veto detector, or depositing less than 50 keV in it.

The fraction of backscattered-coincidence events in the prototype data is somewhat larger (2.8\%) then that predicted by the Monte Carlo (1.8\%). There two possible explanations for this: coincidences with bremstrahlung photons; and electrons that strike the veto detector first and then backscatter into the energy detector. Neither of these phenomena were included in the Monte Carlo but they would add events to the measured coincidence spectrum. 
\par
For an electron kinetic energy of 1 MeV in plastic scintillator, the bremstrahlung yield (energy fraction) is 0.004, 
and the probability of producing a photon greater than 50 keV is approximately 0.016 \cite{Pag72,Koc59}.
Bremstrahlung is emitted predominantly in the forward direction; the probability of emission into the negative hemisphere relative to the electron momentum is only about 20\%. Also the veto detectors are not thick enough to stop all photons in the energy range 0.05--1.0 MeV. Therefore we expect a veto coincidence rate of less than 0.3\% due to bremstrahlung; this is at most a minor contribution to the excess.
\par
In the geometry of our setup (see Figure \ref{F:beamSet}) it is not possible for a primary electron to strike a veto paddle before striking the energy detector, {\em unless} it first penetrates and is redirected by the knife edge aperture. If this happens the electron may strike a veto paddle at a glancing angle and have a moderately high probability of backscattering from there into the energy detector. This certainly does occur at some small but difficult to predict rate; it cannot be prevented;  and these events are indistinguishable in our apparatus from electrons that strike the energy detector first and backscatter into the veto detector. We believe that this is the main contribution to the excess coincidence rate. Of course, it remains possible that ETRAN either overpredicts or underpredicts the actual backscattering probability; we cannot establish the accuracy of the predicted rate from these data alone.
\par
We must also explain the moderately large (1.2\%) low energy tail in the energy detector spectrum that is not associated with coincidence events. 
There are three known sources for such events: (1) electrons that scatter from or penetrate matter in the beam transport system, such as the beam pipe, flanges, or the knife-edge aperture, and lose energy prior to entering the spectrometer; (2) x rays associated with the primary beam or scattered electrons; and (3) electrons that backscatter from the energy detector and do not produce a veto signal in the veto detector. We believe that the majority of these events are due to the first two sources. First, we found that the size and shape of the low energy tail in the energy detector spectrum was very sensitive to beam conditions and the placement of shielding close to the beam. We experimented with many different conditions and setups to limit both the scattered, low-energy electrons in the beam and the x rays that enter the spectrometer. The final configuration represents our best and most successful efforts, but we do not believe we completely eliminated these contributions to the low-energy tail. Second, an independent estimate of the backscatter veto efficiency (see below) showed that the efficiency was probably in the range 87--95\%, so we should not expect source (3) to be a major contributor.
\par 
Because there are sources of low-energy events in the energy detector that are not associated with electron backscatter, we cannot determine the backscatter veto efficiency of the spectrometer simply by comparing the energy detector singles and coincidence spectra. We need an independent way to estimate the veto efficiency. First, consider the data displayed in Figure \ref{F:vdet6scat}, a scatter plot of coincidence events in the energy detector and a single veto paddle (paddle 6). Most of the coincidences fall into two horizontal bands in the veto detector ADC spectrum. The lower band, at about channel 150, are events where a single photoelectron was produced in the veto detector PMT. To ensure that most single photoelectron events were above the discriminator threshold we set the PMT gain (voltage on the base) very high, 2900 V. As a result, most of the higher energy veto events, corresponding to many photoelectrons in the event, are in the veto ADC overflow channel, the upper horizontal band. The broad, nearly vertical band on the left with a negative slope contains backscattered electrons. The slope demonstrates the sharing of the incident energy between the energy and veto detectors. We note that the energy resolution of the veto detector is quite poor, due to relatively poor light collection in this geometry. For a 
given energy detector energy, the associated coincidence veto energy spectrum is very broad. This means that the light collection efficiency of the veto paddle varies significantly as a function of the place where the electron strikes it. Given the shape of the veto paddle and light guides this is not surprising. The backscatter rejection efficiency is therefore closely related to the probability of producing at least one photoelectron in the PMT, averaged over the electron energy and the struck position on the veto paddle.
\begin{figure}
\begin{center}
\vspace{1in}
SEE FIGURE AT END OF FILE\\[1in]
\end{center}
\caption{\label{F:vdet6scat}A scatter plot of energy detector pulse height (ADC) vs. veto detector pulse height (paddle 6 only) for coincidence events. The lowest horizontal band contains random coincidences with noise. The horizontal band just above that contains true coincidences (backscatter events) where the veto detector PMT detected a single photoelectron. The upper horizontal band contains coincidences in the overflow channel of the veto detector ADC. The broad vertical band on the right contains backscattered events where multiple photoelectrons were detected in the veto detector PMT.}
\end{figure}
\par
\begin{figure}
\begin{center}
\includegraphics[width=5.0in]{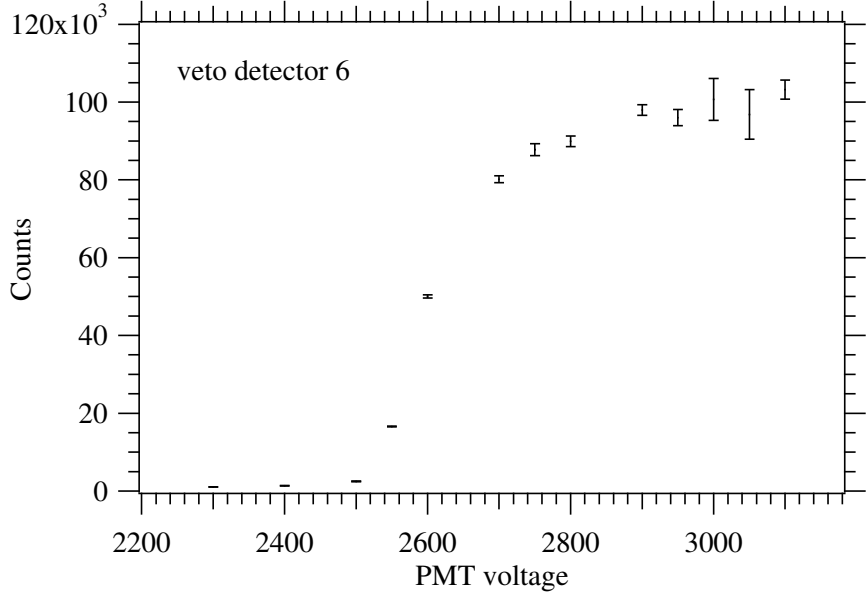}
\end{center}
\caption{\label{F:offline6} Number of single-photoelectron events detected using a visible light source, vs. PMT base voltage, in an offline test.}
\end{figure}

We must address an important question here: What fraction of single photoelectron (p.e.) events will generate a pulse above the discriminator threshold? To answer this question we performed the following ancillary measurement with no electron beam. When the room lights are on in the target room, a small amount of light leaks into the detector chamber through the Kapton window which is slightly transparent, and can reach the veto PMT via small gaps in the mylar wrapping on the paddle. This light produces only single p.e. events in the veto detector. We made equal time measurements of these events with the room lights on and off, to subtract dark current, producing a pure single p.e. spectrum. We then repeated this as a function of PMT base voltage. The results, for veto paddle 6, are shown in Figure \ref{F:offline6}. This curve is effectively the integral of the single p.e. pulse height distribution (approximately a Gaussian). At a PMT voltage of 2500 V the discriminator threshold is above the entire single p.e. distribution. At about 2600 V the threshold is close to the peak of the distribution. At 2900 V the threshold is below the distribution and the single p.e. counting efficiency remains constant as the PMT voltage is increased further. When the electron beam data were taken the veto detector PMT voltage was 2900 V, so the single photoelectron counting efficiency was effectively 100\%. 
\par
\begin{figure}
\begin{center}
\includegraphics[width=5.0in]{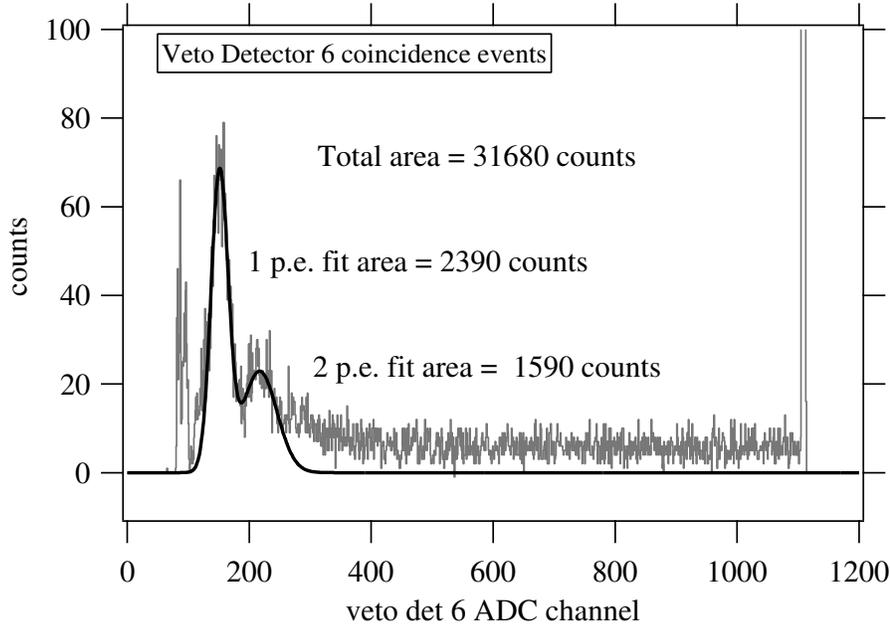}
\end{center}
\caption{\label{F:vdet6ADC} The pulse height (ADC) spectrum of veto detector 6 for events in coincidence with the energy detector. The two fit peaks correspond to one and two photoelectrons detected. Most events are in the overflow channel (channel 1105) which is vertically off-scale.}
\end{figure}

From this we can conclude that a valid veto pulse was generated whenever the backscattered electron produced at least one photoelectron in a veto detector PMT. There is of course some chance that a backscattered electron struck the veto detector and produced no photoelectrons. In that case the event was not vetoed. We must now concentrate on estimating that probability. Figure \ref{F:vdet6ADC} shows a histogram of the pulse height distribution of veto detector 6 for events in coincidence with the energy detector. Most events are in the ADC overflow channel (1105). The single and two photoelectron peaks, and a hint of the three photoelectron peak, are evident on the left. If the first two are fit to a double Gaussian, we obtain 2390 single photoelectron and 1590 two photoelectron events. The total number of events, including overflows, is 31,680. Each scintillation photon produced in the veto paddle has an independent, small chance of being successfully transported to the photomultiplier and producing a photoelectron. Therefore, for a particular electron energy incident on a particular spot on the paddle, we can expect the number of photoelectrons collected to conform to a Poisson distribution:
\begin{equation}
P(\mu,n) = \frac{e^{-\mu} \mu^n}{n!}
\label{E:Poisson}
\end{equation}
in which $P(\mu,n)$ is the probability of producing $n$ photoelectrons when the average is $\mu$. Clearly $\mu$ depends on electron energy and the struck position on the paddle. The pulse height distribution of Figure \ref{F:vdet6ADC}, which is also the distribution of photoelectron number, is the sum of the Poisson distributions for the various electron energies and paddle positions. The average number of photoelectrons in the total distribution is greater than 20, so the one and two photoelectron peaks are dominated by the Poisson distributions associated with the lowest electron energies and the worst light collection. We will make the simplifying assumption these peaks belong to a single Poisson distribution with a single mean $\mu$. From Equation \ref{E:Poisson} we have: $P(1) / P(2) = 2/\mu$ and $P(0) / P(1) = 1/\mu$. Using the fit peak areas in Figure \ref{F:vdet6ADC} we obtain an estimate of 1790 events that produced zero photoelectrons. Now in this process we neglected the three (and higher) photoelectron peaks and the tails of higher peaks that lie under our fit curve, so we probably underestimated the number of zero-PE events. However we believe that, judging from Figure \ref{F:vdet6ADC}, it is fair to say that the true value of the ratio $P(1) / P(2)$ lies in the range 1--3. This gives us an estimated range of 1195--3585 events that produced zero photoelectrons. Comparing this to the total number of coincidence events in 
Figure \ref{F:vdet6ADC} we can thus estimate the overall backscatter veto efficiency to be in the range 90--96\%.
\par
\begin{figure}
\begin{center}
\includegraphics[width=5.0in]{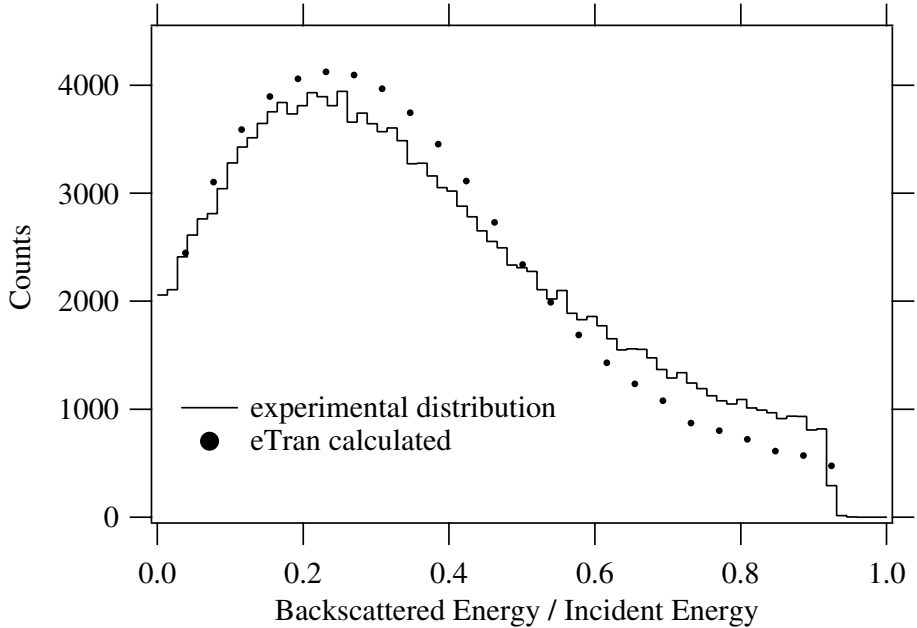}
\end{center}
\caption{\label{F:BSEnDist} The distribution of backscattered energy measured by the missing energy in the energy detector (solid) and the calculated distribution using the ETRAN Monte Carlo on plastic scintillator (dots). The calculated distribution has been convoluted with a Gaussian to simulate for the energy resolution of the detector.}
\end{figure}

Another interesting result that can be extracted from these data is the energy distribution of electrons backscattered from the plastic scintillator energy detector. The net coincidence spectrum (see Figure \ref{F:eDetLog}) shows the distribution of energies that backscattered electrons deposited in the energy detector. The inverse of this spectrum, subtracted from the calculated beam energy of 976 keV, is the spectrum of missing energy. It gives the energy distribution carried by the backscattered electrons. This can be directly compared to the ETRAN simulated result, calculated using 976 keV electrons incident with the same collimation on a thick plastic scintillator. This comparison is presented in Figure \ref{F:BSEnDist}. To facilitate the comparison, the ETRAN spectrum has been convoluted with a Gaussian corresponding to the energy detector energy resolution. Note that the agreement is quite good, although there is some disagreement on the fraction of backscattered electrons that carry more than 50\% of the full energy. This may result from the fact that electrons that backscatter and carry high energy are somewhat more likely to produce at least one p.e. and be vetoed.

\section{Conclusions}
Together, the Monte Carlo simulation of the spectrometer in the conditions of the $a$ coefficient experiment including the realistic magnetic field, and the prototype test run on the electron Van de Graaff, provide a good indication of how the spectrometer will perform in the experiment. The Monte Carlo demonstrated that the low energy tail in the electron response function, caused by electrons that backscatter and deposit less than 75\% of their initial energy in the energy detector, will be less than 0.2\% assuming that all backscattered electrons that deposit more than 50 keV in the veto detector produce a veto signal. The prototype test showed that, due to relatively poor light collection in the veto paddles, the veto efficiency depends more on the position struck by the electron than the energy deposited. It showed that we can expect only 90--96\% of these events to produce a veto signal. Looking at Table \ref{T:MCResults}, we see that this will cause the tail to increase by as much as 10\% of {\sf f$_{\sf 75}$V}. Even with this increase the total low-energy tail for electron energy 300--700 keV will still be less than 0.5\%. Therefore we conclude that this spectrometer will be suitable for the experiment.
\par
We plan to design and build an improved version of the spectrometer for use in the $a$ coefficient experiment. One important improvement will be to bend the veto detector light guides at a smaller angle to increase light collection efficiency, and therefore increase the veto efficiency. It should be possible to do this and still maintain the total length of the spectrometer within the desired maximum of 72 cm. Another improvement will be to shorten the energy detector light guide to increase its light collection efficiency. This should produce a slightly better energy resolution.

\section{Acknowledgements}
We thank Boris Yerozolimsky, Lev Goldin, and Richard Wilson of Harvard University for important discussions related to this project. We are grateful to Jeffrey Nico and David Gilliam of NIST for their interest in this work and helpful comments. This work was supported in part by U.S. Department of Energy Interagency Agreement DE-AI02-93ER40784.


\begin{thebibliography}{00}
\bibitem{NISTdisc} Certain trade names and company products are mentioned in the text or identified in illustrations in order to 
adequately specify the experimental procedure and equipment used. In no case does such identification imply recommendation or endorsement by the National Institute of Standards and Technology, nor does it imply that the products are necessarily the best available for the purpose.
\end{thebibliography}
\end{document}